\documentclass[letter,longauth]{aa} 
\usepackage{graphicx}
\usepackage{hyperref}
\usepackage{natbib}
\bibpunct{(}{)}{;}{a}{}{,}
\usepackage{txfonts}
\usepackage{xspace}

\newcommand{\jybm}{{\ensuremath{\rm Jy\,beam^{-1}}}\xspace}

\newcommand{\co}{{\ensuremath{{\rm ^{12}CO}}}\xspace}

\newcommand{\msun}{{\ensuremath{M_{\odot}}}\xspace}
\newcommand{\msunyr}{{\ensuremath{M_{\odot}\,{\rm yr^{-1}}}}\xspace}
\newcommand{\lsun}{{\ensuremath{L_{\odot}}}\xspace}
\newcommand{\kms}{{\ensuremath{{\rm km\, s^{-1}}}}\xspace}

\newcommand{\kmsarcsec}{{\ensuremath{\rm km\, s^{-1}\, arcsec^{-1}}}\xspace}

\begin{document}
\title{The first ALMA view of IRAS~16293-2422
\thanks{Continuum and spectral data are  available in electronic form at the CDS via anonymous ftp to
\url{cdsarc.u-strasbg.fr} (130.79.128.5) or via \url{http://cdsweb.u-strasbg.fr/cgi-bin/qcat?J/A+A/}}}
\titlerunning{First ALMA view of IRAS~16293}
   \subtitle{Direct detection of infall onto source B
   \\and high-resolution kinematics of source A}

\author{Jaime E. Pineda\inst{1,2} 
\and 
Ana\"elle J. Maury\inst{1}
\and
Gary A. Fuller\inst{2} 
\and
Leonardo Testi\inst{1,3} 
\and
Diego Garc\'{\i}a-Appadoo\inst{4,5} 
\and
Alison B. Peck\inst{5,6} 
\and
Eric Villard\inst{5} 
\and
Stuartt A. Corder\inst{6}
\and
Tim A. van~Kempen\inst{5,7}
\and
Jean L. Turner\inst{8}
\and
Kengo Tachihara\inst{5,9}
\and 
William Dent\inst{5} 
}

\authorrunning{J. E. Pineda et al.}

\institute{European Southern Observatory (ESO), Garching, Germany
\email{jaime.pineda@manchester.ac.uk}
\and
UK ARC Node, Jodrell Bank Centre for Astrophysics, School of Physics and Astronomy, University of Manchester, Manchester, M13 9PL, UK
\and
INAF-Osservatorio Astrofisico di Arcetri, Largo E. Fermi 5, I-50125 Firenze, Italy
\and
European Southern Observatory, Alonso de C\'ordova 3107, Vitacura, Casilla 19001, Santiago 19, Chile
\and
Joint ALMA Observatory, Alonso de C\'ordova 3107, Vitacura, Santiago, Chile
\and
North American ALMA Science Center, National Radio Astronomy Observatory, 520 Edgemont Road, Charlottesville, VA 22903, USA
\and
Leiden Observatory, Leiden University, PO Box 9513, 2300 RA Leiden, The Netherlands
\and
Department of Physics and Astronomy, UCLA, Los Angeles, CA 90095, USA
\and
National Astronomical Observatory of Japan, Chile Observatory, 2-21-1 Osawa Mitaka Tokyo 181-8588 Japan
}

\date{June 22, 2012; accepted in A\&A}

\abstract%
{}
{We focus on the kinematical properties of a proto-binary to study the infall and 
rotation of gas towards its two protostellar components.}
{We present ALMA Science Verification observations with 
high-spectral resolution of IRAS 16293-2422 at 220.2\, GHz. 
The wealth of molecular lines in this source and the very high spectral resolution offered by ALMA allow 
us to study the gas kinematics with unprecedented detail.
}
{We present the first detection of an inverse P-Cygni profile toward source B in the three brightest lines. 
The line profiles are fitted with a simple two-layer model to derive an infall rate of $4.5\times 10^{-5}$\msunyr. 
This infall detection would rule-out the previously suggested possibility that source B is a T Tauri star.
A position velocity diagram for source A shows evidence of rotation with an axis close to the line-of-sight.
}
{} 

\keywords{ISM: clouds -- stars: formation  -- ISM: molecules -- 
ISM: individual (IRAS\,16293-2422)}

\maketitle

\section{Introduction}
IRAS 16293-2422 (hereafter I16293) is a well-studied Class 0 protostar with a bolometric luminosity of 32\ \lsun , embedded in a 3\ \msun envelope of size $\sim$3000\ AU \citep{Correia_2004-IRAS16293_ISO_SCUBA_SED}. It is located in the nearby 
$\rho$ Ophiuchi star-forming region, at a distance of $\sim$120\ pc \citep{Knude_1998-HIPPARCOS_distances, Loinard_2008-Oph_Distance}.
I16293 has been shown to consist of two main sources denoted as components A and B (hereafter I16293A and I16293B), separated by 5$\arcsec$ (600\ AU) in the plane of the sky \citep{Looney_2000-BIMA_Cont_Survey}. 
The structure of I16293A might be more complex than that of I16293B: 
two centimeter sources  \citep[A1 and A2;][]{Wooteen_1989-IRAS16293_Protobinary} and 
two submillimeter sources \citep[Aa and Ab;][]{Chandler_2005-IRAS16293_highres} were detected toward I16293A, 
while I16293B has not shown any sign of substructure at these wavelengths so far.

Despite its low luminosity, I16293 has a rich chemistry, with hot-core-like (hot-corino) properties at scales of $\sim$100 AU 
\citep[e.g.,][]{Blake_1994-IRAS16293_Si_S_Chemistry, Ceccarelli_1998-IRAS16293_hotcore_D2CO, Schoier_2002-IRAS16293_HotCore, Cazaux_2003-HotCore_IRAS16293, Caux_2011-TIMASS_Observations, Coutens_2012-IRAS16293_D2O}.
Single-dish observations and modeling 
\citep{vanDishoeck_1995-IRAS16293_Chemistry, Ceccarelli_2000-IRAS16293_H2CO, Schoier_2002-IRAS16293_HotCore} 
suggest that the emission from organic molecules arises from small-scale regions toward the continuum sources, 
where the temperature ($\sim$80--100~K) would allow grain-mantle evaporation. 
Moreover, recent interferometric observations have shown that some complex species are more abundant 
toward I16293A than toward I16293B 
\citep{Bottinelli_2004-IRAS16293_HotCorino_PdBI, Kuan_2004-IRAS16293_SMA_2HotCores}.

Early interferometric studies \citep{Bottinelli_2004-IRAS16293_HotCorino_PdBI,Kuan_2004-IRAS16293_SMA_2HotCores} 
have shown that the spectra toward I16293A have broad lines (FWHMs up to 8 \kms), 
whereas the lines toward I16293B are much narrower (typically less than 2 \kms wide). 
The determination of the centroid velocity toward each source is a matter of debate, where different studies 
present different results. 
While \cite{Bisschop_2008-IRAS16293_SMA_Molecules} argued that the high-excitation lines of the complex organics peak at 
$V_{LSR}$ =1.5--2.5 \kms toward both sources, \cite{Jorgensen_2011-SMA_IRAS16293} 
reported average systemic velocities of 3.2 and 2.7 \kms and average line widths of 2.6 and 1.9 \kms for A and B, respectively.

Several studies \cite[e.g.,][]{Kuan_2004-IRAS16293_SMA_2HotCores,Bottinelli_2004-IRAS16293_HotCorino_PdBI, Rao_2009-IRAS16293_SMA_Magnetic, Bisschop_2008-IRAS16293_SMA_Molecules, Crimier_2010-IRAS16293_Structure,Jorgensen_2011-SMA_IRAS16293} have focused on I16293 to characterize the respective nature of the two continuum sources. 
Indeed, I16293 is known to have a quadrupolar outflow \citep{Walker_1988-IRAS16293_Outflow, Mizuno_1990-IRAS16293_Outflow,Castets_2001-IRAS16293_shocks, Rao_2009-IRAS16293_SMA_Magnetic} suggesting the existence of a binary system. However, significant physical differences (chemical and kinematic) exist, suggesting different natures or different evolutionary stages of the two continuum sources A and B. 
Whereas it is generally agreed that the I16293A component is protostellar in nature, it has been suggested that the I16293B component either represented a more evolved (T Tauri) star \citep{Stark_2004-IRAS16293_Infall_HDO, Takakuwa_2007-IRAS16293_SMA_HCN} or alternatively a very young object \citep{Chandler_2005-IRAS16293_highres}.

Evidence for large-scale infall toward this protostellar system has been found using single-dish observations 
\citep{Walker_1986-IRAS16293_Infall_SingleDish, Narayanan_1998-IRAS16293_Infall}. 
\cite{Chandler_2005-IRAS16293_highres} suggested that there is evidence for infall toward 
I16293B from the tentative 
absorption feature seen in SO when imaged using only baselines longer than 55k$\lambda$.

\section{Data}
I16293 was observed on August 16-18, 2011 in ALMA band 6 as part of the ALMA Science Verification effort, 
yielding a total on-source observing time of 5.4 hours.
The array was used to point successively at two different positions, 
($\alpha,\delta$)$_{\rm J2000}$=(16:32:22.99,-24:28:36.100) and (16:32:22.71,-24:28:32.326), allowing the observation 
of two overlapping fields and obtaining an homogeneous response over the full extent of the binary 
system from the primary beam response of the ALMA 12-m antennas.
The spectral setup consisted of observing the H$_{2}$CCO $11_{(1,11)}$--$10_{(1,10)}$ line at 220.178\,GHz, 
using one baseband centered on 220.182~GHz, with a channel width of 61~kHz and bandwidth of 234.375~MHz. 
Atmospheric variations at each antenna were monitored continuously using water vapor radiometers (WVRs), in addition to regular observations of a nearby phase reference source.  
Gain changes were tracked using regular hot/ambient load measurements

The uncalibrated dataset was released, together with spectrally averaged, low-resolution calibrated images,
and is publicly available since April 2012\footnote{http://almascience.eso.org/almadata/sciver/IRAS16293Band6/}. 
We calibrated and imaged the original interferometric data with the CASA\footnote{http://casa.nrao.edu/} software version 3.3.0. 
Absolute flux calibration was performed using the CASA Butler-JPL-Horizons 2010 model applied to Neptune observations, which results in an estimated flux uncertainty of $\pm$10--15\%.
Bandpass calibration was performed using observations of the strong quasar J1924-292, while time-dependent gain calibration was derived by regularly observing (each 12~minutes) the nearby quasar J1625-254.
The calibrated measurement set was spectrally binned to a channel spacing of 120~kHz 
(Hanning smoothing, 0.16\,\kms resolution), and then CLEANed using the Clark algorithm.
The use of 16 antennas with a longest baseline of 220-m results in maps with a synthesized beam 
size 2.2\arcsec$\times$1.0\arcsec at 220\,GHz. 
The continuum image was produced by using line-free channels between 220.103 and 220.112\,GHz, 
extracted from the visibilities table and imaged separately. 
After a first imaging of the continuum map, we performed self-calibration on the continuum data which 
lowered the rms 
noise in the continuum map by a factor of three. 
The resulting continuum map has a final rms noise level of 3.6~m\jybm with a 1.9\arcsec$\times$0.9\arcsec beam 
(Fig.\ref{fig-map}), while the theoretical rms in this map should have been $\sim$0.7~m\jybm, 
if the noise had been completely decorrelated over the 175 channels used to build the continuum map. 
The continuum was removed from the visibilities to produce the spectral cube, and we applied the self-calibration 
derived from the continuum emission to the spectral data. 
The resulting rms in the line-free channels of the spectral cube is $\sim$4.5~m\jybm per 120~kHz-channel, 
while the theoretical noise is 4.3~m\jybm.

These ALMA data have a spectral coverage included in the SMA observations of \cite{Jorgensen_2011-SMA_IRAS16293},
but the ALMA FWHM-beam is almost a factor 2 lower, while the spectral resolution is increased by a factor 5 and the sensitivity by a factor 15-20.

\section{Results}

\subsection{Continuum emission and molecular complexity}

\begin{figure*}
\includegraphics[width=17cm]{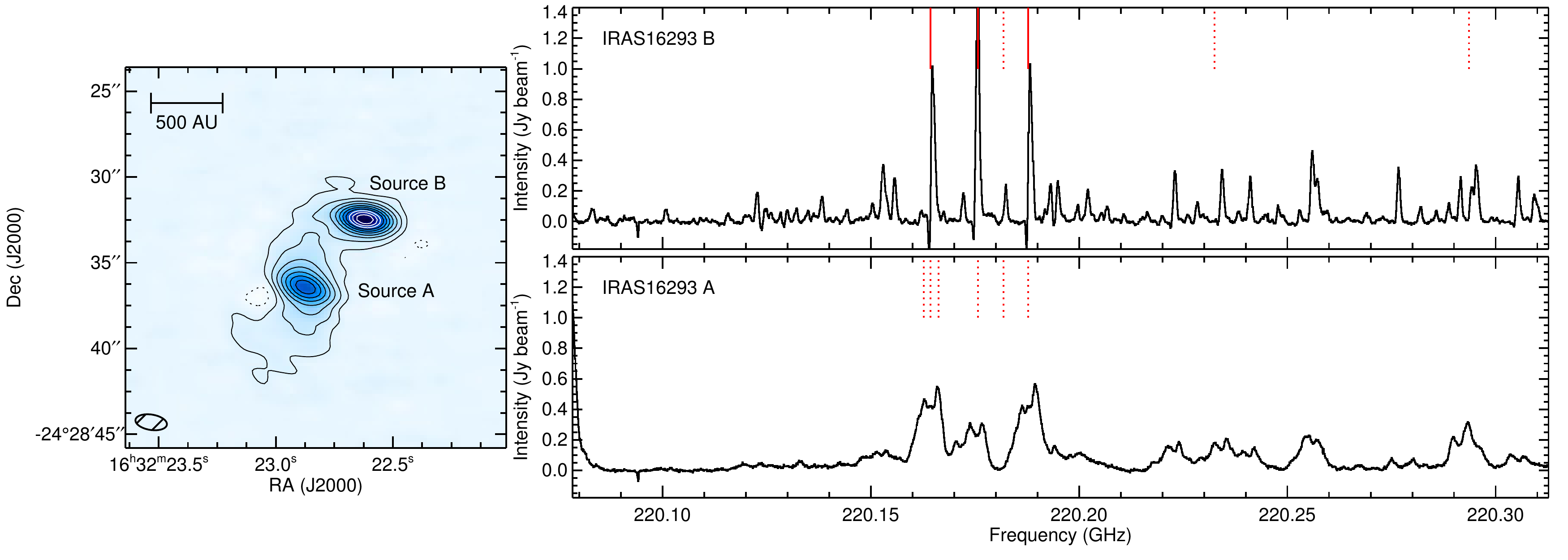}
\caption{\emph{Left:} 
Continuum map of I16293 obtained at 220.1\,GHz with ALMA. The synthesized  beam 
is shown in the bottom left corner.
The rms noise is 3.6~m\jybm in the central 20$\arcsec$ region of the map.
Contours are drawn to -3, 3, 10, 22, 39, 61, 88, 120, and 157 time the rms noise, where 
negative contours are plotted using dotted lines. 
The peak continuum flux densities for sources A and B are 0.539$\pm$0.004 and 1.066$\pm$0.004\,\jybm, respectively.
\label{fig-map}
\emph{Right:} 
Spectra of sources A and B are shown in the bottom and top panels, respectively. 
The rms noise in the line-free channels is 4.5 m\jybm.
The red lines (solid and dotted) show the molecular transitions previously identified by \cite{Jorgensen_2011-SMA_IRAS16293}, 
and the solid lines mark the transition where the inverse P-Cygni profile is found.
\label{fig-spec}}
\end{figure*}

Sources I16293A and I16293B are clearly detected in the continuum map, shown in Fig.~\ref{fig-map}. 
The continuum peak fluxes of A and B agree with the fluxes reported previously by 
\citet{Bottinelli_2004-IRAS16293_HotCorino_PdBI} at 230~GHz with a similar beam. 
The ALMA observations also trace some extended continuum emission connecting sources A and B, 
which was not detected in the PdBI map of \citet{Bottinelli_2004-IRAS16293_HotCorino_PdBI}, 
and is most likely due to spatial filtering and poorer sensitivity of the PdBI observations.
This extended continuum emission traces the inner region of the common protostellar envelope as 
traced by single-dish observations, 
and includes the position at which \cite{Remijan_2006-IRAS16293_Infall_Disk} 
detected an inverse P-Cygni profile in the low-density tracer \co(2--1).

The continuum emission is used to estimate the mass as 
\[
M_{1.3~mm} = 1.3~\msun \left(\frac{F_{1.3~mm}}{1{\rm Jy}}\right) \left(\frac{d}{200~{\rm pc}}\right)^{2}
\left( e^{0.36\,(30~K/T)} -1\right)~,
\]
where we assumed optically thin emission and a dust opacity per dust mass ($\kappa_{1.3~mm}$) 
of 0.86\,cm$^{2}$\,g$^{-1}$ \citep[thick ice mantles coagulated at $10^{5}$\,cm$^{-3}$ from][]{OH94} 
and a gas-to-dust ratio of 100. 
The total flux, measured over the 10-$\sigma$ region, 
for sources A and B is 1.30$\pm$0.11 and 1.06$\pm$0.15~Jy, respectively, 
which implies a mass of 0.21 and 0.26\,\msun for sources A and B, respectively.

Thanks to the improved sensitivity and spectral resolution offered by ALMA, many ($\sim$50) molecular transitions 
are detected toward the two continuum sources, while previous SMA observations at this frequency only detected 
six molecular lines toward each object \citep{Jorgensen_2011-SMA_IRAS16293}. 
The peak spectra toward source A and B are shown in Fig.~\ref{fig-spec}, where we can appreciate the difference in
the line profile toward these two objects. 
A thorough line identification and comparison of chemical properties will be addressed in a future publication.


\subsection{Evidence for Infalling gas toward source B}

\begin{figure}
\includegraphics[width=8.6cm]{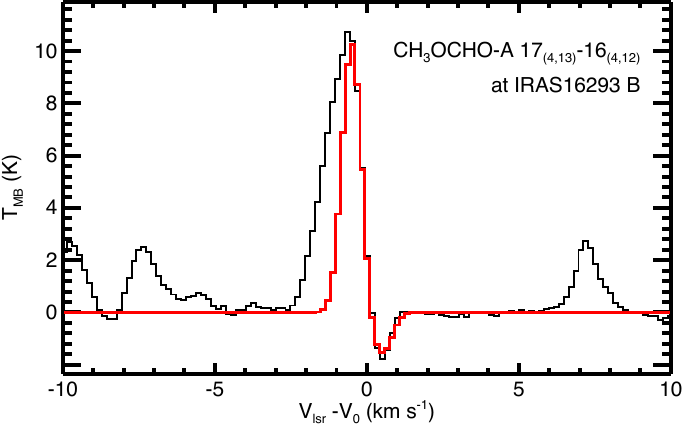}
\caption{Spectrum toward the continuum peak of I16293B for CH$_{3}$OCHO-A. 
The red line shows the best two-layer model of infall fit. 
The best-fit model parameters are listed in Table~\ref{table-fit}. 
See Fig.~\ref{fig-infall-all} for fits of all three molecules. 
\label{fig-infall}}
\end{figure}

Toward I16293B, inverse P-Cygni profiles are unambiguously detected in three 
molecular lines (see Fig.\ref{fig-infall} and \ref{fig-infall-all}). 
These line profiles are clear evidence for infall in source B. 
Although an absorption feature in SO ($7_{6}$-$6_{6}$) toward source B was reported by 
\cite{Chandler_2005-IRAS16293_highres} as 
suggestive of infall, the ALMA observations here analyzed are the \emph{first observations} of the 
inverse P-Cygni profile toward this source.

The inverse P-Cygni line profiles are modeled to extract the velocity information of the infalling gas 
and estimate the infall rates. 
Here we used the simple two-slab model described by \cite{Myers_1996-Line_Profile_Model_Infall}
with the modification introduced by \cite{DiFrancesco_2001-NGC1333IRAS4_dynamics} to take into 
account the continuum source. 
The model fits the infall velocity of the layers, $V_{in}$, its optical depth, $\tau_{0}$, velocity dispersion, $\sigma_{v}$, 
and excitation temperature of the layer on the rear, $T_{r}$, while the excitation temperature of the foreground 
layer, $T_{f}$, is fixed at 3~K; details of the modeling procedure are described in more detail in Appendix~\ref{app-model}.

The CH$_{3}$OCHO-A molecular line with an inverse P-Cygni profile is shown in Fig.~\ref{fig-infall}, 
with the best two-layer model shown in red 
(all three molecular lines with an inverse P-Cygni profile are shown in Fig.~\ref{fig-infall-all}). 
The blue excess in the line profiles is not fitted, and therefore the fit is mostly constrained by the absorption feature, 
see \cite{DiFrancesco_2001-NGC1333IRAS4_dynamics,Kristensen_2012-WISH_Infall_Outflow_Profiles} for 
similar procedures. 
Despite its simplicity, the two-layer model contains more parameters than can be fully constrained 
because of degeneracies in the model, but allows a robust determination of the 
velocity information. 
The parameters of the best two-layer model are listed in Table~\ref{table-fit}. 

\begin{table}
\caption{Parameters for fitted ``two-layer'' model\label{table-fit}}
\centering
\begin{tabular}{lcccc}
\hline\hline
 Line & $T_r$  &$\tau_0$ &$V_{in}$  &$\sigma_v$\\
      & (K)    &         & (\kms)       & (\kms)\\
\hline
CH$_3$OCHO-E & 44$\pm$3   &  0.48$\pm$0.04 & 0.49$\pm$0.02 & 0.30$\pm$0.02\\
CH$_3$OCHO-A & 46$\pm$3   &  0.45$\pm$0.04 & 0.49$\pm$0.02 & 0.31$\pm$0.01 \\
H$_2$CCO &  60$\pm$10   &  0.33$\pm$0.05 & 0.51$\pm$0.07 & 0.39$\pm$0.03\\
\hline
\end{tabular}
\tablefoot{
$V_{in}$ is the infall velocity of the layers. 
$\tau_{0}$ is the peak optical depth of each layer.
$\sigma_{v}$ is the layer velocity dispersion. 
$T_{r}$ is the rear layer excitation temperature. 
$T_{f}$ is the front layer excitation temperature. 
The following parameters are kept fixed: 
$V_{LSR}=3.4\,\kms$, $T_c=20$~K, $T_f=3$~K, and $\Phi=0.3$. 
See Appendix~\ref{app-model} for a description.
}
\end{table}

The infall rate is estimated assuming spherical symmetry as 
$\dot{M}_{infall}= 4\pi r_{in}^{2}\, n_{in}\, \mu m_{\rm H}\, V_{in},$
where at radius $r_{in}$ the infall velocity and density have values of  $V_{in}$ and $n_{in}$, 
respectively; and $\mu$ is 
mean molecular weight of the gas (2.3).

The infall radius, $r_{in}$, can be estimated assuming that the infall velocity is only free-fall, 
$M = V_{in}^{2}r_{in}/2G=4\pi r_{in}^{3}\, n_{in}\, \mu m_{\rm H}/3,$
while also estimating the central mass, $M$, as that from a uniform density sphere with radius $r_{in}$ and 
density $n_{in}$.
Therefore, the accretion rate is estimated as 
\begin{equation}
\dot{M}_{infall}= \frac{3 V_{in}^{3}}{2G} = 4.5 \times 10^{-5} \left(\frac{V_{in}}{0.5\,\kms}\right)^{3}\msunyr~.
\end{equation}
The infall rates obtained using the best fits are 4.2, 4.5, and 4.8$\times 10^{-5}$\,\msunyr 
for CH$_3$OCHO-E, CH$_3$OCHO-A, and H$_2$CCO, respectively.

\subsection{Velocity maps}

\begin{figure}
\resizebox{\hsize}{!}{\includegraphics{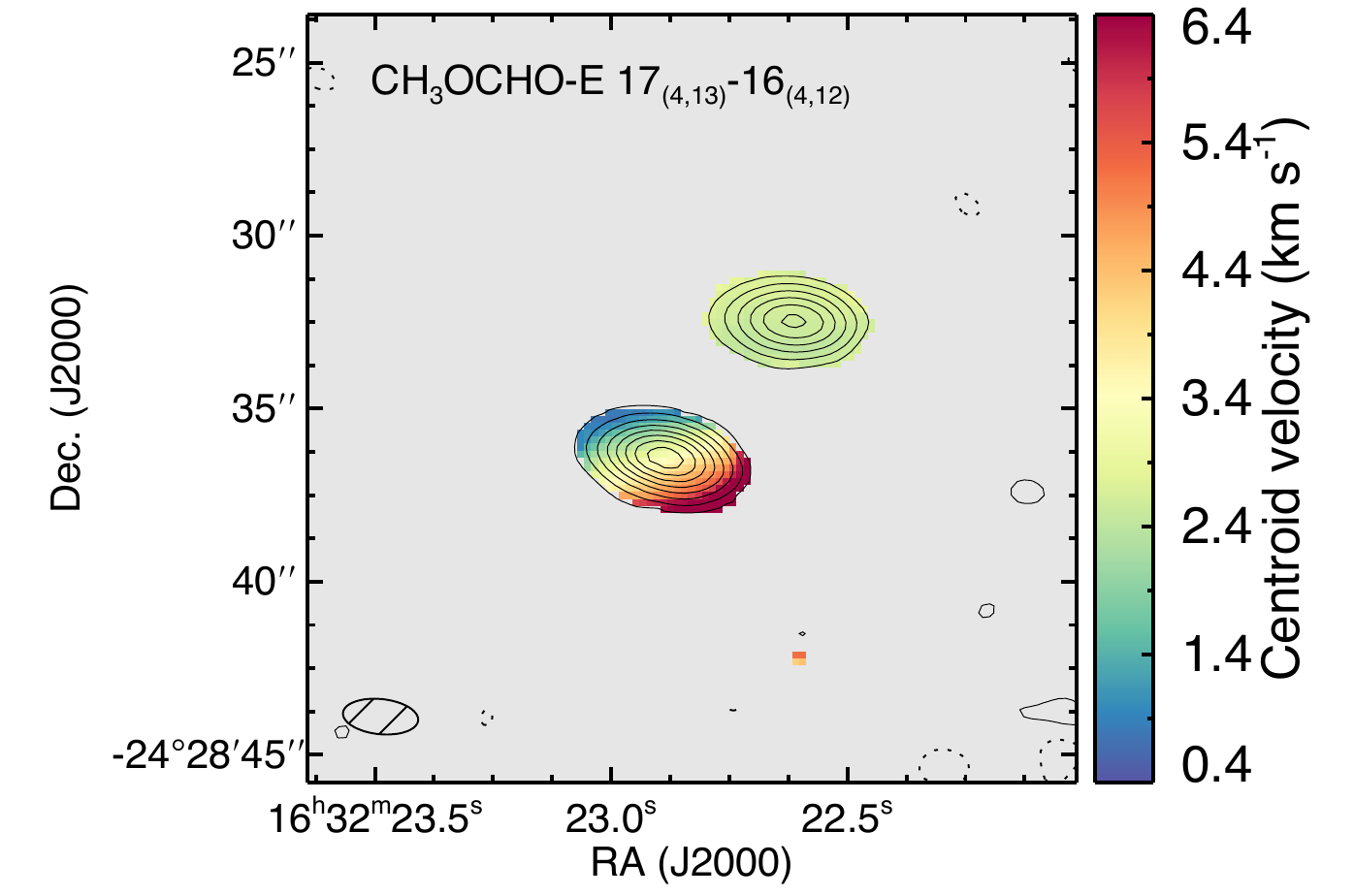}}
\caption{Intensity-weighted velocity for CH$_3$OCHO-E.
Contours are drawn to -3, 3, 10, 22, 39, 61, 88, 120, and 157 times the rms noise of the integrated intensity, 
15m\jybm\kms, where negative contours are plotted using dotted lines.
The beam size is shown at the bottom left corner. 
See Fig.~\ref{fig-w-line} for velocity maps of all three molecules. 
\label{fig-w-line-single}
}
\end{figure}

The intensity-weighted velocity map for CH$_3$OCHO-E is shown in Fig.~\ref{fig-w-line-single}, 
where the line-integrated intensity is also overlaid in contours 
(see Fig.~\ref{fig-w-line} for maps of all three molecules). 
The morphology of the emission and its intensity-weighted velocity for CH$_3$OCHO-A and CH$_3$OCHO-E 
are quite similar, which is not surprising given the almost identical energy level and Einstein coefficients 
(see Table~\ref{table-line-spec}). 
For H$_{2}$CCO, which has a lower energy level than CH$_3$OCHO-A/E, 
the emission is more extended, but with a similar velocity structure. 
Source A presents a prominent velocity gradient, with a velocity variation of $\sim$5.5\,\kms across 
the extent of the molecular emission, which is aligned along the axis connecting the submillimeter 
sources Aa and Ab previously detected by \cite{Chandler_2005-IRAS16293_highres}.
In source B  there is no evidence for such a strong velocity gradient, and in fact the velocity variation  
across the entire source B is $\sim$0.4\,\kms, which is more than an order of magnitude lower than in source A.

The position velocity (PV) diagrams for source A (along the direction shown in Fig.~\ref{fig-w-line}) 
are consistent with rotation of a disk, which is close to being edge-on (to explain the 
strong velocity variation). 
The PV diagrams for the three molecular lines used are shown in Fig.~\ref{fig-pv-line}.  
This orientation would also agree with the orientation of the outflow seen in SiO by \cite{Rao_2009-IRAS16293_SMA_Magnetic}. 
However, the present ALMA data do not rule-out the possibility of two velocity components at scales smaller 
than our synthesized beam, due for example to the emission of several sources in a multiple system 
\citep[e.g. Aa, Ab;][]{Chandler_2005-IRAS16293_highres}.

\section{Discussion and conclusion}

The molecules where the P-Cygni profile is observed are complex organics, usually associated with 
the hot-corino and not with the ambient cloud \cite[e.g.][]{Caselli_1993-Orion_Chemistry} 
although some of them have also been detected toward outflows \citep{Arce_2008-Complex_Molecules_Outflow_L1157}.
Therefore, we can rule out the absorption from the ambient cloud or ambient envelope, and associate the 
observed P-Cygni profiles to infall from the inner envelope. 
This infall detection would rule-out the possibility that source B is a young T Tauri star, as previously suggested by 
\cite{Stark_2004-IRAS16293_Infall_HDO}.
We notice that it is only thanks to the improvement in spectral resolution and sensitivity 
of the presented ALMA observations 
compared to previous interferometric observations of this source 
that the inverse P-Cygni profile could  be detected and modeled.

The infall velocity, $V_{in}$, derived from the three different molecular lines is the same (0.50$\pm 0.01\,\kms$)
and supersonic. 
If the velocity dispersion of the absorbing layer is only caused by thermal motions ($c_s=0.32\,\kms$ for gas at 30~K), 
the infall velocity is between 1.3 and 1.5$\times$ the sound speed,  i.e. supersonic infall.
The infall velocity and derived infall rates are consistent with the values derived 
by \cite{Kristensen_2012-WISH_Infall_Outflow_Profiles} for a small sample of low-mass protostars. 
It is important to note that for source A we could not identify an inverse P-Cygni profile, if present, 
because of the steep velocity gradient. 

The PV diagram for source A is consistent with rotation seen almost edge-on and generated by 
a central object of 0.53\msun, which is consistent with previous estimations using the source kinematics 
\citep[e.g.][]{Chandler_2005-IRAS16293_highres}. 
However, this mass estimate is almost twice as high as the mass estimated using 
continuum emission \citep[see also][]{Rao_2009-IRAS16293_SMA_Magnetic}. 
This discrepancy can be reconciled if the dust emission is partially optically thick at submillimeter wavelengths. 

The lack of a stronger velocity gradient in source B might be due to either rotation happening at  
smaller scales than those probed in these observations or because source B is closer to being face-on, 
which has previously been suggested based on the analysis of the continuum emission 
\citep{Rodriguez_2005-IRAS16293_B_edge_on_disk,Loinard_2007-IRAS16293_Structure}.

This work illustrates the new capabilities of ALMA, which open a new window for studying the kinematics 
and chemistry in the early stages of star-formation. 
Using ALMA to carry out observations of the kinematics of source A at 
higher angular resolution will for example allow one to distinguish between the two possible scenarios we propose here 
to explain the velocity gradient detected in the present observations, e.g edge-on disk or close binary system.

\begin{acknowledgements}
This paper makes use of the following ALMA data: 
ADS/JAO.ALMA\#2011.0.00007.SV . ALMA is a partnership of ESO (representing 
its member states), NSF (USA) and NINS (Japan), together with NRC 
(Canada) and NSC and ASIAA (Taiwan), in cooperation with the Republic of 
Chile. The Joint ALMA Observatory is operated by ESO, AUI/NRAO and NAOJ.
JEP and AA has received funding from the European CommunityÕs 
Seventh Framework Programme (/FP7/2007-2013/) under grant agreement No 229517.
\end{acknowledgements}

\Online
\begin{appendix}
\section{The two-layer model}\label{app-model}
Each inverse P-Cygni profile was fitted using the simple two-slab model described by 
\cite{Myers_1996-Line_Profile_Model_Infall}
with the modification introduced by \cite{DiFrancesco_2001-NGC1333IRAS4_dynamics} to take into 
account the continuum source, see also \cite{Kristensen_2012-WISH_Infall_Outflow_Profiles}. 
The model assumes that the central continuum source is optically thick, emitting as a blackbody of temperature $T_{c}$, 
and filling a fraction of the beam, $\Phi$. 
Two layers of gas, front and rear, are infalling toward the central source with an infall velocity and velocity dispersion 
$V_{in}$ and $\sigma_{v}$, respectively. 
The rear layer is illuminated by the background radiation, $T_{b}$. 
If for each layer the peak optical depth is $\tau_{0}$, then the expected line emission at velocity $V$ can be expressed as 
\begin{eqnarray}
\Delta T_{B}&=&
\left(J_{f}-J_{cr}\right) \left[1-e^{-\tau_{f}}\right] \nonumber \\
&&+\left(1-\Phi\right) (J_{r}-J_{b})  \left[1-e^{-(\tau_{r}+\tau_{f})}\right]~,
\end{eqnarray}
where 
\begin{equation}
J_{cr}= \Phi\,J_{c}+ (1-\Phi)\,J_{r}~,
\end{equation}
and 
\begin{eqnarray}
\tau_{f}&=&\tau_{0}\, \exp{\left[\frac{-(V-(V_{LSR}+V_{in}))^{2}}{2\sigma_{v}^{2}}\right]}\,\\
\tau_{r}&=&\tau_{0}\, \exp{\left[\frac{-(V-(V_{LSR}-V_{in}))^{2}}{2\sigma_{v}^{2}}\right]}\,,
\end{eqnarray}
where 
$V_{LSR}$ is the source velocity, 
and the radiation temperature is defined as 
\begin{equation}
J_{x}= \frac{T_{0}}{[\exp{(T_{0}/T_{x})}-1]}~,
\end{equation}
where $T_{0}\equiv h\nu_{0}/k_{B}$, and $\nu_{0}$ is the line rest frequency.

Since the front absorbing layer is probably subthermally excited, we assumed a conservative value for the 
excitation front layer of $T_{f}=3$~K.
The continuum temperature, $T_{c}$, was chosen such that the continuum radiation temperature, 
$J_{c}(T_{c})$, matches the peak continuum flux in the image.
The continuum source beam filling fraction, $\Phi$, which is not well constrained, was set to 0.3 to use 
the same value as 
\cite{DiFrancesco_2001-NGC1333IRAS4_dynamics}, who modeled 
interferometric observations with a similar angular resolution as we did; this value is consistent 
with the line modeling results of coarse angular resolution Herschel data \cite{Kristensen_2012-WISH_Infall_Outflow_Profiles}. 
Although the value of $\Phi$ used here is arbitrary, 
we have also performed line profile fits with higher values of $\Phi$ and the obtained infall velocity and 
velocity dispersion are unchanged, while the values for $T_{r}$ and $\tau_{0}$ are increased and decreased, respectively.
Similarly, if a higher value of  $T_{c}$ is used, $T_{r}$ and $\tau_{0}$ are increased and decreased, respectively, 
while we obtain the same results for $V_{in}$ and $\sigma_{v}$. 
Therefore our results are robust to the value of $\Phi$ and $T_{c}$ used. 
An initial set of models with the centroid velocity, $V_{LSR}$, as a free parameter were run, and all gave a $V_{LSR}$ of 
3.4$\pm$0.1\,\kms. Given these results, we decided to fix the value of $V_{LSR}$ at 3.4\,\kms and  
reduce the number of free variables in the fit.

The $\chi^{2}$ was minimized using the \verb+mpfitfun+ procedure \citep{IDL_MPFIT}, where a 
spectrum is generated for different parameters and then compared to the observed line profile. 
The following parameters were kept fixed during the minimization: 
$V_{LSR}$, $T_c$, $T_f$, and $\Phi$. 

\end{appendix}

\begin{appendix}
\section{Tables}
Table~\ref{obj-pos} lists the position of the continuum sources. 
Table~\ref{table-line-spec} lists the information for the spectral lines analyzed. 

\begin{table}[h]
\caption{Position of sources\label{obj-pos}}
\begin{tabular}{lcc}
\hline\hline
Source &    R.A. (J2000)&    Decl. (J2000)\\
 & (hh:mm:ss) & (dd:mm:ss) \\
\hline
IRAS16293~A  & 16:32:22.873    & -24:28:36.50\\
IRAS16293~B  & 16:32:22.623    & -24:28:32.49\\
\hline
\end{tabular}
\end{table}

\begin{table}[h]
\caption{Spectral parameters for lines with inverted P-Cygni profile\label{table-line-spec}}
\begin{tabular}{lccccc}
\hline\hline
Molecule &  Transition & Frequency\tablefootmark{a} & $E_{u}$\tablefootmark{b} &Ref.\\
&      & (GHz)                      & (K)                      &\\
\hline
CH$_3$OCHO-E& $17_{(4,13)}-16_{(4,12)}$& 220.166888 & 103.15  & 1 \\
CH$_3$OCHO-A& $17_{(4,13)}-16_{(4,12)}$& 220.190285 & 103.14  & 1 \\
H$_2$CCO          & $11_{(1,11)}-10_{(1,10)}$& 220.17757   &    76.48  & 2 \\
\hline
\end{tabular}
\tablebib{The spectroscopic references given are the most recent cited in the CDMS and JPL databases 
\citep{Pickett_1998-JPL_Line_Catalogue,Muller_2001-CDMS,Muller_2005-CDMS}. 
(1)~\citet{Maeda_2008-CH3OCHO_param_v2}, 
(2)~\citet{Guarnieri_2003-H2CCO_param_v2}}
\end{table}

\end{appendix}

\begin{appendix}
\section{Inverse P-Cygni profiles, Centroid velocity  and position velocity maps}
\begin{figure*}
\includegraphics[width=8.6cm]{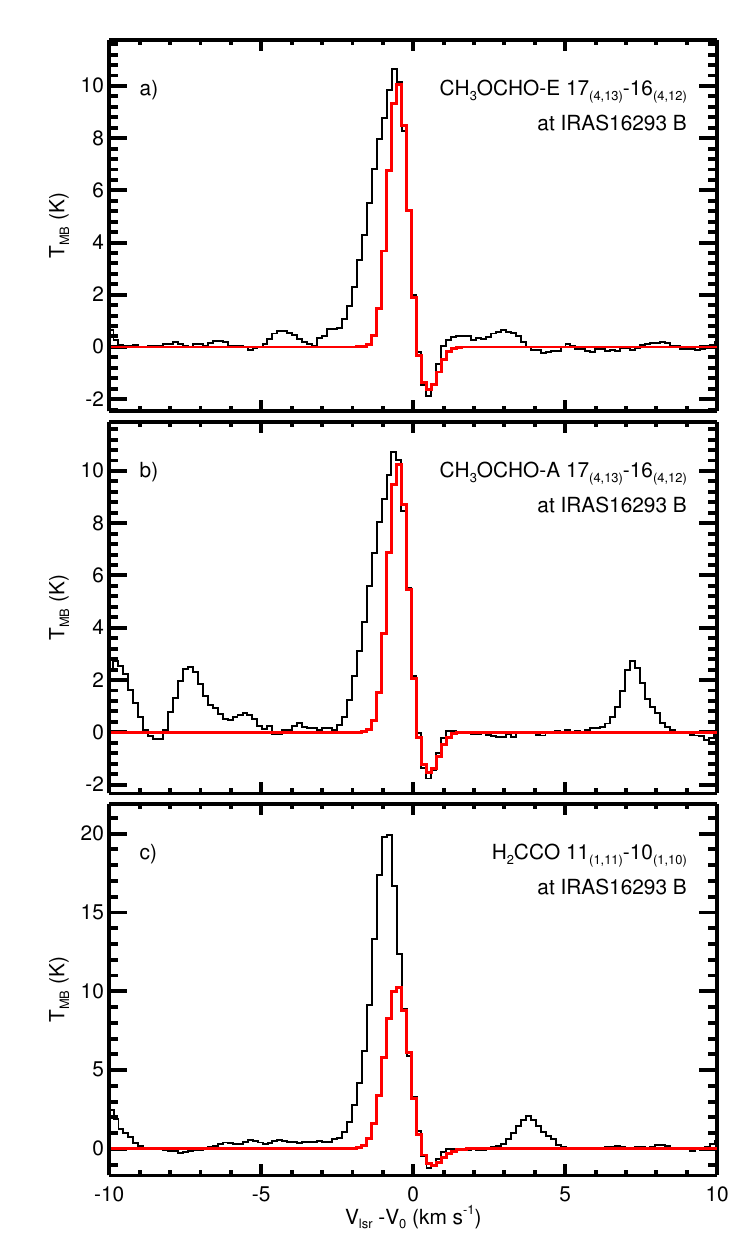}
\caption{Molecular line spectra from the ALMA data toward the continuum peak of I16293B.
In each panel, a red line shows the best two-layer model of infall fit for each spectrum. 
The best-fit model parameters are listed in Table~\ref{table-fit}.
\label{fig-infall-all}}
\end{figure*}

\begin{figure*}
\includegraphics[width=17cm]{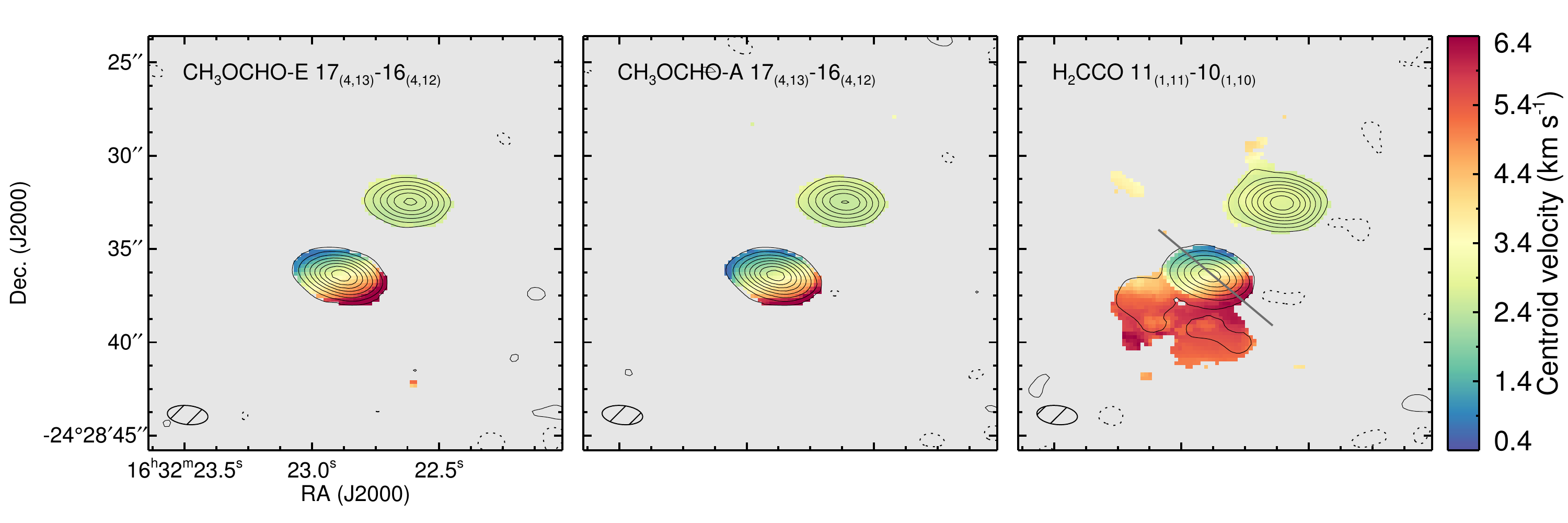}
\caption{Intensity-weighted velocity for the three lines studied, CH$_{3}$OCHO-A (left panel)
CH$_{3}$OCHO-E (middle panel), and H$_{2}$CCO (right panel). 
Contours are drawn to -3, 3, 10, 22, 39, 61, 88, 120, and 157 times the rms noise of the integrated intensity, 
15m\jybm\kms, where negative contours are plotted using dotted lines.
The beam size is shown at the bottom left corner. 
The gray line is the cut used for the position velocity diagram shown in Fig.~\ref{fig-pv-line}.
\label{fig-w-line}
}
\end{figure*}

\begin{figure*}
\includegraphics[width=17cm]{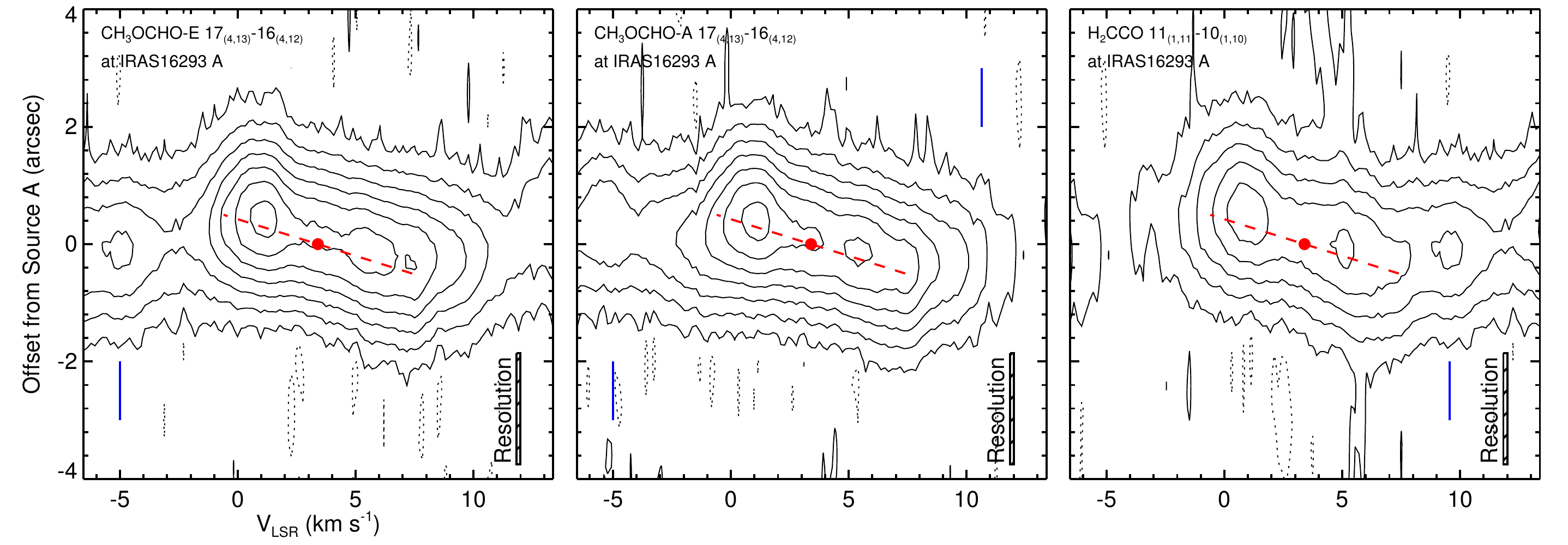}
\caption{Position velocity maps of source A, along the direction shown in the rightmost panel of Fig.~\ref{fig-w-line}, 
for the three lines studied. 
From left to right: CH$_{3}$OCHO-A, CH$_{3}$OCHO-E, and H$_{2}$CCO. 
Contours are drawn to -3, 3, 10, 22, 39, 61, 88, 120, and 157 times the rms noise, 
4.5m\jybm, where negative contours are plotted using dotted lines. 
The red dashed line and red solid circle show the 
8\,\kmsarcsec velocity gradient and position of source A ($V_{LSR}$=3.4\kms), respectively. 
Notice that in all panels an adjacent molecular transition line is marked with a vertical 
blue line, which can also be identified in Fig.~\ref{fig-infall}. 
The spatial resolution is shown at the bottom right corner. 
\label{fig-pv-line}
}
\end{figure*}

In this appendix, we show the inverse P-Cygni profiles and their respective models in Fig.~\ref{fig-infall-all}. 
The centroid velocity maps for all three molecules studied, 
CH$_{3}$OCHO-A/E and H$_{2}$CCO, are shown in Fig.~\ref{fig-w-line}. 
The position velocity diagrams along the cut shown in Fig.~\ref{fig-w-line} are shown in Fig.~\ref{fig-pv-line}

\end{appendix}


\begin{thebibliography}{42}
\expandafter\ifx\csname natexlab\endcsname\relax\def\natexlab#1{#1}\fi

\bibitem[{{Arce} {et~al.}(2008){Arce}, {Santiago-Garc{\'{\i}}a},
  {J{\o}rgensen}, {Tafalla}, \&
  {Bachiller}}]{Arce_2008-Complex_Molecules_Outflow_L1157}
{Arce}, H.~G., {Santiago-Garc{\'{\i}}a}, J., {J{\o}rgensen}, J.~K., {Tafalla},
  M., \& {Bachiller}, R. 2008, \apjl, 681, L21

\bibitem[{{Bisschop} {et~al.}(2008){Bisschop}, {J{\o}rgensen}, {Bourke},
  {Bottinelli}, \& {van Dishoeck}}]{Bisschop_2008-IRAS16293_SMA_Molecules}
{Bisschop}, S.~E., {J{\o}rgensen}, J.~K., {Bourke}, T.~L., {Bottinelli}, S., \&
  {van Dishoeck}, E.~F. 2008, \aap, 488, 959

\bibitem[{{Blake} {et~al.}(1994){Blake}, {van Dishoeck}, {Jansen}, {Groesbeck},
  \& {Mundy}}]{Blake_1994-IRAS16293_Si_S_Chemistry}
{Blake}, G.~A., {van Dishoeck}, E.~F., {Jansen}, D.~J., {Groesbeck}, T.~D., \&
  {Mundy}, L.~G. 1994, \apj, 428, 680

\bibitem[{{Bottinelli} {et~al.}(2004){Bottinelli}, {Ceccarelli}, {Neri},
  {Williams}, {Caux}, {Cazaux}, {Lefloch}, {Maret}, \&
  {Tielens}}]{Bottinelli_2004-IRAS16293_HotCorino_PdBI}
{Bottinelli}, S., {Ceccarelli}, C., {Neri}, R., {et~al.} 2004, \apjl, 617, L69

\bibitem[{{Caselli} {et~al.}(1993){Caselli}, {Hasegawa}, \&
  {Herbst}}]{Caselli_1993-Orion_Chemistry}
{Caselli}, P., {Hasegawa}, T.~I., \& {Herbst}, E. 1993, \apj, 408, 548

\bibitem[{{Castets} {et~al.}(2001){Castets}, {Ceccarelli}, {Loinard}, {Caux},
  \& {Lefloch}}]{Castets_2001-IRAS16293_shocks}
{Castets}, A., {Ceccarelli}, C., {Loinard}, L., {Caux}, E., \& {Lefloch}, B.
  2001, \aap, 375, 40

\bibitem[{{Caux} {et~al.}(2011){Caux}, {Kahane}, {Castets}, {Coutens},
  {Ceccarelli}, {Bacmann}, {Bisschop}, {Bottinelli}, {Comito}, {Helmich},
  {Lefloch}, {Parise}, {Schilke}, {Tielens}, {van Dishoeck}, {Vastel},
  {Wakelam}, \& {Walters}}]{Caux_2011-TIMASS_Observations}
{Caux}, E., {Kahane}, C., {Castets}, A., {et~al.} 2011, \aap, 532, A23

\bibitem[{{Cazaux} {et~al.}(2003){Cazaux}, {Tielens}, {Ceccarelli}, {Castets},
  {Wakelam}, {Caux}, {Parise}, \& {Teyssier}}]{Cazaux_2003-HotCore_IRAS16293}
{Cazaux}, S., {Tielens}, A.~G.~G.~M., {Ceccarelli}, C., {et~al.} 2003, \apjl,
  593, L51

\bibitem[{{Ceccarelli} {et~al.}(1998){Ceccarelli}, {Castets}, {Loinard},
  {Caux}, \& {Tielens}}]{Ceccarelli_1998-IRAS16293_hotcore_D2CO}
{Ceccarelli}, C., {Castets}, A., {Loinard}, L., {Caux}, E., \& {Tielens},
  A.~G.~G.~M. 1998, \aap, 338, L43

\bibitem[{{Ceccarelli} {et~al.}(2000){Ceccarelli}, {Loinard}, {Castets},
  {Tielens}, \& {Caux}}]{Ceccarelli_2000-IRAS16293_H2CO}
{Ceccarelli}, C., {Loinard}, L., {Castets}, A., {Tielens}, A.~G.~G.~M., \&
  {Caux}, E. 2000, \aap, 357, L9

\bibitem[{{Chandler} {et~al.}(2005){Chandler}, {Brogan}, {Shirley}, \&
  {Loinard}}]{Chandler_2005-IRAS16293_highres}
{Chandler}, C.~J., {Brogan}, C.~L., {Shirley}, Y.~L., \& {Loinard}, L. 2005,
  \apj, 632, 371

\bibitem[{{Correia} {et~al.}(2004){Correia}, {Griffin}, \&
  {Saraceno}}]{Correia_2004-IRAS16293_ISO_SCUBA_SED}
{Correia}, J.~C., {Griffin}, M., \& {Saraceno}, P. 2004, \aap, 418, 607

\bibitem[{{Coutens} {et~al.}(2012){Coutens}, {Vastel}, {Caux}, {Ceccarelli},
  {Bottinelli}, {Wiesenfeld}, {Faure}, {Scribano}, \&
  {Kahane}}]{Coutens_2012-IRAS16293_D2O}
{Coutens}, A., {Vastel}, C., {Caux}, E., {et~al.} 2012, \aap, 539, A132

\bibitem[{{Crimier} {et~al.}(2010){Crimier}, {Ceccarelli}, {Maret},
  {Bottinelli}, {Caux}, {Kahane}, {Lis}, \&
  {Olofsson}}]{Crimier_2010-IRAS16293_Structure}
{Crimier}, N., {Ceccarelli}, C., {Maret}, S., {et~al.} 2010, \aap, 519, A65

\bibitem[{{Di Francesco} {et~al.}(2001){Di Francesco}, {Myers}, {Wilner},
  {Ohashi}, \& {Mardones}}]{DiFrancesco_2001-NGC1333IRAS4_dynamics}
{Di Francesco}, J., {Myers}, P.~C., {Wilner}, D.~J., {Ohashi}, N., \&
  {Mardones}, D. 2001, \apj, 562, 770

\bibitem[{{Guarnieri} \& {Huckauf}(2003)}]{Guarnieri_2003-H2CCO_param_v2}
{Guarnieri}, A. \& {Huckauf}, A. 2003, Z. Naturforsch, 58, 275

\bibitem[{{J{\o}rgensen} {et~al.}(2011){J{\o}rgensen}, {Bourke}, {Nguyen
  Luong}, \& {Takakuwa}}]{Jorgensen_2011-SMA_IRAS16293}
{J{\o}rgensen}, J.~K., {Bourke}, T.~L., {Nguyen Luong}, Q., \& {Takakuwa}, S.
  2011, \aap, 534, A100

\bibitem[{{Knude} \& {Hog}(1998)}]{Knude_1998-HIPPARCOS_distances}
{Knude}, J. \& {Hog}, E. 1998, \aap, 338, 897

\bibitem[{{Kristensen} {et~al.}(2012){Kristensen}, {van Dishoeck}, {Bergin},
  {Visser}, {Y{\i}ld{\i}z}, {San Jose-Garcia}, {J{\o}rgensen}, {Herczeg},
  {Johnstone}, {Wampfler}, {Benz}, {Bruderer}, {Cabrit}, {Caselli}, {Doty},
  {Harsono}, {Herpin}, {Hogerheijde}, {Karska}, {van Kempen}, {Liseau},
  {Nisini}, {Tafalla}, {van der Tak}, \&
  {Wyrowski}}]{Kristensen_2012-WISH_Infall_Outflow_Profiles}
{Kristensen}, L.~E., {van Dishoeck}, E.~F., {Bergin}, E.~A., {et~al.} 2012,
  \aap, 542, A8

\bibitem[{{Kuan} {et~al.}(2004){Kuan}, {Huang}, {Charnley}, {Hirano},
  {Takakuwa}, {Wilner}, {Liu}, {Ohashi}, {Bourke}, {Qi}, \&
  {Zhang}}]{Kuan_2004-IRAS16293_SMA_2HotCores}
{Kuan}, Y.-J., {Huang}, H.-C., {Charnley}, S.~B., {et~al.} 2004, \apjl, 616,
  L27

\bibitem[{{Loinard} {et~al.}(2007){Loinard}, {Chandler}, {Rodr{\'{\i}}guez},
  {D'Alessio}, {Brogan}, {Wilner}, \& {Ho}}]{Loinard_2007-IRAS16293_Structure}
{Loinard}, L., {Chandler}, C.~J., {Rodr{\'{\i}}guez}, L.~F., {et~al.} 2007,
  \apj, 670, 1353

\bibitem[{{Loinard} {et~al.}(2008){Loinard}, {Torres}, {Mioduszewski}, \&
  {Rodr{\'{\i}}guez}}]{Loinard_2008-Oph_Distance}
{Loinard}, L., {Torres}, R.~M., {Mioduszewski}, A.~J., \& {Rodr{\'{\i}}guez},
  L.~F. 2008, \apjl, 675, L29

\bibitem[{{Looney} {et~al.}(2000){Looney}, {Mundy}, \&
  {Welch}}]{Looney_2000-BIMA_Cont_Survey}
{Looney}, L.~W., {Mundy}, L.~G., \& {Welch}, W.~J. 2000, \apj, 529, 477

\bibitem[{{Maeda} {et~al.}(2008){Maeda}, {De Lucia}, \&
  {Herbst}}]{Maeda_2008-CH3OCHO_param_v2}
{Maeda}, A., {De Lucia}, F.~C., \& {Herbst}, E. 2008, Journal of Molecular
  Spectroscopy, 251, 293

\bibitem[{{Markwardt}(2009)}]{IDL_MPFIT}
{Markwardt}, C.~B. 2009, in Astronomical Society of the Pacific Conference
  Series, Vol. 411, Astronomical Data Analysis Software and Systems XVIII, ed.
  {D.~A.~Bohlender, D.~Durand, \& P.~Dowler}, 251--+

\bibitem[{{Mizuno} {et~al.}(1990){Mizuno}, {Fukui}, {Iwata}, {Nozawa}, \&
  {Takano}}]{Mizuno_1990-IRAS16293_Outflow}
{Mizuno}, A., {Fukui}, Y., {Iwata}, T., {Nozawa}, S., \& {Takano}, T. 1990,
  \apj, 356, 184

\bibitem[{{M{\"u}ller} {et~al.}(2005){M{\"u}ller}, {Schl{\"o}der}, {Stutzki},
  \& {Winnewisser}}]{Muller_2005-CDMS}
{M{\"u}ller}, H.~S.~P., {Schl{\"o}der}, F., {Stutzki}, J., \& {Winnewisser}, G.
  2005, Journal of Molecular Structure, 742, 215

\bibitem[{{M{\"u}ller} {et~al.}(2001){M{\"u}ller}, {Thorwirth}, {Roth}, \&
  {Winnewisser}}]{Muller_2001-CDMS}
{M{\"u}ller}, H.~S.~P., {Thorwirth}, S., {Roth}, D.~A., \& {Winnewisser}, G.
  2001, \aap, 370, L49

\bibitem[{{Myers} {et~al.}(1996){Myers}, {Mardones}, {Tafalla}, {Williams}, \&
  {Wilner}}]{Myers_1996-Line_Profile_Model_Infall}
{Myers}, P.~C., {Mardones}, D., {Tafalla}, M., {Williams}, J.~P., \& {Wilner},
  D.~J. 1996, \apjl, 465, L133+

\bibitem[{{Narayanan} {et~al.}(1998){Narayanan}, {Walker}, \&
  {Buckley}}]{Narayanan_1998-IRAS16293_Infall}
{Narayanan}, G., {Walker}, C.~K., \& {Buckley}, H.~D. 1998, \apj, 496, 292

\bibitem[{{Ossenkopf} \& {Henning}(1994)}]{OH94}
{Ossenkopf}, V. \& {Henning}, T. 1994, \aap, 291, 943

\bibitem[{{Pickett} {et~al.}(1998){Pickett}, {Poynter}, {Cohen}, {Delitsky},
  {Pearson}, \& {M{\"u}ller}}]{Pickett_1998-JPL_Line_Catalogue}
{Pickett}, H.~M., {Poynter}, R.~L., {Cohen}, E.~A., {et~al.} 1998, \jqsrt, 60,
  883

\bibitem[{{Rao} {et~al.}(2009){Rao}, {Girart}, {Marrone}, {Lai}, \&
  {Schnee}}]{Rao_2009-IRAS16293_SMA_Magnetic}
{Rao}, R., {Girart}, J.~M., {Marrone}, D.~P., {Lai}, S.-P., \& {Schnee}, S.
  2009, \apj, 707, 921

\bibitem[{{Remijan} \& {Hollis}(2006)}]{Remijan_2006-IRAS16293_Infall_Disk}
{Remijan}, A.~J. \& {Hollis}, J.~M. 2006, \apj, 640, 842

\bibitem[{{Rodr{\'{\i}}guez} {et~al.}(2005){Rodr{\'{\i}}guez}, {Loinard},
  {D'Alessio}, {Wilner}, \& {Ho}}]{Rodriguez_2005-IRAS16293_B_edge_on_disk}
{Rodr{\'{\i}}guez}, L.~F., {Loinard}, L., {D'Alessio}, P., {Wilner}, D.~J., \&
  {Ho}, P.~T.~P. 2005, \apjl, 621, L133

\bibitem[{{Sch{\"o}ier} {et~al.}(2002){Sch{\"o}ier}, {J{\o}rgensen}, {van
  Dishoeck}, \& {Blake}}]{Schoier_2002-IRAS16293_HotCore}
{Sch{\"o}ier}, F.~L., {J{\o}rgensen}, J.~K., {van Dishoeck}, E.~F., \& {Blake},
  G.~A. 2002, \aap, 390, 1001

\bibitem[{{Stark} {et~al.}(2004){Stark}, {Sandell}, {Beck}, {Hogerheijde}, {van
  Dishoeck}, {van der Wal}, {van der Tak}, {Sch{\"a}fer}, {Melnick}, {Ashby},
  \& {de Lange}}]{Stark_2004-IRAS16293_Infall_HDO}
{Stark}, R., {Sandell}, G., {Beck}, S.~C., {et~al.} 2004, \apj, 608, 341

\bibitem[{{Takakuwa} {et~al.}(2007){Takakuwa}, {Ohashi}, {Bourke}, {Hirano},
  {Ho}, {J{\o}rgensen}, {Kuan}, {Wilner}, \&
  {Yeh}}]{Takakuwa_2007-IRAS16293_SMA_HCN}
{Takakuwa}, S., {Ohashi}, N., {Bourke}, T.~L., {et~al.} 2007, \apj, 662, 431

\bibitem[{{van Dishoeck} {et~al.}(1995){van Dishoeck}, {Blake}, {Jansen}, \&
  {Groesbeck}}]{vanDishoeck_1995-IRAS16293_Chemistry}
{van Dishoeck}, E.~F., {Blake}, G.~A., {Jansen}, D.~J., \& {Groesbeck}, T.~D.
  1995, \apj, 447, 760

\bibitem[{{Walker} {et~al.}(1986){Walker}, {Lada}, {Young}, {Maloney}, \&
  {Wilking}}]{Walker_1986-IRAS16293_Infall_SingleDish}
{Walker}, C.~K., {Lada}, C.~J., {Young}, E.~T., {Maloney}, P.~R., \& {Wilking},
  B.~A. 1986, \apjl, 309, L47

\bibitem[{{Walker} {et~al.}(1988){Walker}, {Lada}, {Young}, \&
  {Margulis}}]{Walker_1988-IRAS16293_Outflow}
{Walker}, C.~K., {Lada}, C.~J., {Young}, E.~T., \& {Margulis}, M. 1988, \apj,
  332, 335

\bibitem[{{Wootten}(1989)}]{Wooteen_1989-IRAS16293_Protobinary}
{Wootten}, A. 1989, \apj, 337, 858

\end{thebibliography}
\end{document}